\documentclass[twocolumn, tighten]{aastex631}
\usepackage{lineno}

\usepackage{nicefrac}
\usepackage{microtype}
\usepackage{booktabs}
\usepackage{rotating}
\usepackage{xcolor}

\newcommand{\kms}{${\rm km\,s}^{-1}$}

\received{?}
\revised{?}
\accepted{?}
\submitjournal{ApJ}
\shorttitle{CGM nebulae around luminous quasars at intermediate redshift}
\shortauthors{Johnson et al.}

\begin{document}

\title{Discovery of optically emitting circumgalactic nebulae around the majority of UV-luminous quasars at intermediate redshift}

\author[0000-0001-9487-8583]{Sean D. Johnson}
\affil{Department of Astronomy, University of Michigan, 1085 S. University, Ann Arbor, MI 48109, USA}
\correspondingauthor{Sean D. Johnson}
\email{seanjoh@umich.edu}

\author[0000-0002-2662-9363]{Zhuoqi (Will) Liu}
\affil{Department of Astronomy, University of Michigan, 1085 S. University, Ann Arbor, MI 48109, USA}

\author[0000-0002-0311-2812]{Jennifer I-Hsiu Li}
\affil{Department of Astronomy, University of Michigan, 1085 S. University, Ann Arbor, MI 48109, USA}
\affil{Michigan Institute for Data Science, University of Michigan, Ann Arbor, MI 48109, USA}

\author[0000-0002-0668-5560]{Joop Schaye}
\affil{Leiden Observatory, Leiden University, PO Box 9513, NL-2300 RA Leiden, The Netherlands																										}
\author[0000-0002-5612-3427]{Jenny E. Greene}
\affil{Department of Astrophysical Sciences, Princeton University, 4 Ivy Lane, Princeton, NJ 08544, USA}

\author[0000-0001-5804-1428]{Sebastiano Cantalupo}
\affil{Department of Physics, University of Milan Bicocca, Piazza della Scienza 3, I-20126 Milano, Italy}

\author[0000-0002-8459-5413]{Gwen C. Rudie}
\affil{The Observatories of the Carnegie Institution for Science, 813 Santa Barbara Street, Pasadena, CA 91101, USA}

\author[0000-0002-2941-646X]{Zhijie Qu}
\affil{Department of Astronomy \& Astrophysics, The University of Chicago, Chicago, IL 60637, USA}

\author[0000-0001-8813-4182]{Hsiao-Wen Chen}
\affil{Department of Astronomy \& Astrophysics, The University of Chicago, Chicago, IL 60637, USA}

\author[0000-0002-9946-4731]{Marc Rafelski}
\affil{Space Telescope Science Institute, 3700 San Martin Drive, Baltimore, MD 21218, USA}
\affil{Department of Physics and Astronomy, Johns Hopkins University, Baltimore, MD 21218, USA}

\author[0000-0003-3938-8762]{Sowgat Muzahid}
\affil{Inter-University Centre for Astronomy and Astrophysics (IUCAA), Post Bag 4, Ganeshkhind, Pune 411007, India}

\author[0000-0002-8739-3163]{Mandy C. Chen}
\affil{Department of Astronomy \& Astrophysics, The University of Chicago, Chicago, IL 60637, USA}

\author{Thierry Contini}
\affil{Institut de Recherche en Astrophysique et Plan\'etologie (IRAP), Universit\'e de Toulouse, CNRS, UPS, F–31400 Toulouse, France}

\author[0000-0002-0417-1494]{Wolfram Kollatschny}
\affil{Institut für Astrophysik und Geophysik, Universität Göttingen, Friedrich-Hund Platz 1, D-37077 Göttingen, Germany}

\author[0000-0002-9141-9792]{Nishant Mishra}
\affil{Department of Astronomy, University of Michigan, 1085 S. University, Ann Arbor, MI 48109, USA}

\author{Patrick Petitjean}
\affil{Institut d'Astrophysique de Paris 98bis Boulevard Arago, 75014, Paris, France}

\author[0000-0002-1690-3488]{Michael Rauch}
\affil{The Observatories of the Carnegie Institution for Science, 813 Santa Barbara Street, Pasadena, CA 91101, USA}

\author[0000-0001-7869-2551]{Fakhri S. Zahedy}
\affil{The Observatories of the Carnegie Institution for Science, 813 Santa Barbara Street, Pasadena, CA 91101, USA}

\begin{abstract}

{We report the discovery of large ionized, [O\,II] emitting circumgalactic nebulae around the majority of thirty UV luminous quasars at $z=0.4-1.4$ observed with deep, wide-field integral field spectroscopy (IFS) with the Multi-Unit Spectroscopy Explorer (MUSE) by the Cosmic Ultraviolet Baryon Survey (CUBS) and MUSE Quasar Blind Emitters Survey (MUSEQuBES). Among the 30 quasars, seven (23\%) exhibit [O\,II] emitting nebulae with major axis sizes greater than 100 kpc, twenty greater than 50 kpc (67\%), and 27 (90\%) greater than 20 kpc. Such large, optically emitting nebulae indicate that cool, dense, and metal-enriched circumgalactic gas is common in the halos of luminous quasars at intermediate redshift. Several of the largest nebulae exhibit morphologies that suggest interaction-related origins. We detect no correlation between the sizes and cosmological dimming corrected surface brightnesses of the nebulae and quasar redshift, luminosity, black hole mass, or radio-loudness, {but find a tentative correlation between the nebulae and rest-frame [O\,II] equivalent width in the quasar spectra. This potential trend suggests a relationship between  ISM content and gas reservoirs on CGM scales.} The [O\,II]-emitting nebulae around the $z\approx1$ quasars are smaller and less common than Ly$\alpha$ nebulae around $z\approx3$ quasars. These smaller sizes can be explained if the outer regions of the Ly$\alpha$ halos arise from scattering in more neutral gas, by evolution in the cool CGM content of quasar host halos, by lower-than-expected metallicities on $\gtrsim50$ kpc scales around $z\approx1$ quasars, or by changes in quasar episodic lifetimes between $z=3$ and $1$.}

\end{abstract}


\section{Introduction}
\label{section:introduction}

Surveys of the star formation rates and available interstellar gas reservoirs
\citep[][]{Kennicutt:2012, Tacconi:2013} as well as the metallicity distribution of stars
\citep[][]{Hayden:2015, Greener:2021} in the Milky Way and distant, massive galaxies
are inconsistent with simple ``closed-box'' models \citep[][]{Tinsley:1974} of galaxy evolution. These
discrepancies can be resolved if galaxies accrete fresh material from external
reservoirs as they evolve \citep[for a review, see][]{Putman:2017}. At the same time,
outflows driven by supernovae \citep[e.g.][]{Dekel:1986} and Active Galactic Nuclei \citep[e.g.][]{Silk:1998}
are required in state-of-the-art cosmological simulations to reproduce the observed stellar masses of
galaxies \citep[e.g.][]{Vogelsberger:2014, Schaye:2015}. These same feedback mechanisms
eject heavy elements produced by stars and supernovae out of the interstellar medium
and into intergalactic space to reproduce the mass-metallicity relation \citep[e.g.][]{Tremonti:2004, Ma:2016}.
Consequently, improving our empirical understanding of the rich set of physical processes that
regulate gas exchange between galaxies and their surrounding cosmic ecosystems
\citep[for a reviews, see][]{Donahue:2022, Faucher-Giguere:2023} represents a key priority of extragalactic
astrophysics \citep[][]{National-Academies-of-Sciences:2021}.

Direct observations of the intergalactic and circumgalactic medium (IGM/CGM) are
critical to developing a more complete understanding of galaxy evolution.
However, the diffuse nature and cool-warm temperature range ($T\approx 10^4-10^6$ K)
of the IGM/CGM around typical galaxies make emission observations difficult
with current facilities {except in the very local Universe \citep[e.g.,][]{Chynoweth:2008, Lokhorst:2022}
and rare cases with unusually high surface brightness. Consequently, emission observations of the IGM/CGM often rely on
deep, coadded observations averaging over hundreds} to thousands of galaxies \citep[e.g.][]{Wisotzki:2016, Dutta:2023, Guo:2023a, Guo:2023}.

Much of our knowledge of the physical nature and extent of the CGM relies on sensitive absorption
spectroscopy of UV bright background sightlines that pass through the halos of foreground galaxies
\citep[for a review, see][]{Tumlinson:2017}. These observations enable constraints on the densities, temperatures,
and metallicities of the CGM around a wide variety of galaxies and across cosmic time
\citep[e.g.,][]{Zahedy:2019, Rudie:2019, Cooper:2021, Zahedy:2021, Qu:2022}. However, the lack of morphological information in absorption-line
studies of the IGM/CGM {necessitates non-trivial model assumptions when attributing the gas to inflows, outflows, or other origins
\citep[see][]{Gauthier:2012, Ho:2017, Schroetter:2019, Zabl:2019} except in rare cases where multiple sightlines enable velocity shear measurements
\citep[e.g.,][]{Chen:2014, Zahedy:2016, Lopez:2018, Lopez:2020}.}

While direct emission studies of the CGM and IGM are not generally feasible for individual galaxies beyond the local Universe,
IGM/CGM emission can be detected in some {rare} or extreme cases. In particular, systems like quasars that
produce a significantly elevated ionizing radiation background increase the equilibrium
recombination rate and temperature of the surrounding IGM/CGM, resulting in increased
emission in recombination and collisionally excited lines \citep[e.g.,][]{Chelouche:2008}.
Consequently, observers have invested time on large ground-based telescopes
toward identifying and following up systems where the IGM/CGM is expected to be observable in emission
with state-of-the-art wide-field Integral Field Spectrographs \citep[IFS;][]{Bacon:2010, Martin:2010, Mateo:2022}.
Despite cosmological surface brightness dimming, the first successful surveys of individual
CGM emission were conducted at $z=2-4$ in H\,I Ly$\alpha$ emission due to the inherent strength of
the line and its redshifting into the optical window. These surveys demonstrated that giant, $\approx 100$ kpc
CGM nebulae are nearly ubiquitous around luminous quasars at $z=2-4$
\citep[][]{Borisova:2016, Cai:2019, OSullivan:2020, Fossati:2021, Mackenzie:2021},
and some extend to hundreds of kpc in projection \citep[e.g.][]{Cantalupo:2014, Cai:2018}.

Recent IFS-enabled discoveries at lower redshifts of $z<1.5$ revealed the potential for studies of $\approx 30-100$ kpc-scale CGM
nebulae observed in emission in non-resonant, rest-frame optical lines (e.g., [O\,II], H$\beta$, and [O\,III]) and resonant NUV lines (e.g., Mg\,II)
around quasars \citep[][]{Johnson:2018, Helton:2021, Johnson:2022, Dutta:2023, Epinat:2023, Liu:2024},
galaxy groups \citep[][]{Epinat:2018, Chen:2019, Leclercq:2022, Dutta:2023, Epinat:2023},
and galaxies {with evidence of recent bursts of star formation} \citep[][]{Rupke:2019, Burchett:2021, Zabl:2021}.
The joint morphological and kinematic analysis of the nebulae combined with deep galaxy surveys enabled by IFS
provide direct insights into the origins of the gas, including interactions, accretion, and outflows.
Furthermore, statistical analysis of velocity-structure functions around the largest of these nebulae provide
new insights into halo-scale turbulence 
\citep[][]{Chen:2023, Chen:2023b}. However, while such giant, relatively high surface brightness optically emitting
nebulae have led to significant insights in individual case studies, their incidence rate, and properties around quasars
have yet to be quantified in statistical samples.

Here, we report the first IFS survey constraining the incidence of optically emitting circumgalactic nebulae
around 30 UV-luminous, unobscured quasars at $z=0.4-1.4$ using deep observations from the
Multi Unit Spectroscopic Explorer \citep[MUSE;][]{Bacon:2010} collected by the MUSE Quasar Blind Emitters Survey \citep[MUSEQuBES; e.g., ][]{Dutta:2023a} and the
Cosmic Ultraviolet Baryon Surveys \citep[CUBS;][]{Chen:2020}. The paper proceeds as follows:
In section \ref{section:observations}, we describe the observations and data reduction steps.
In section \ref{section:quasars}, we characterize the properties of the MUSEQuBES and CUBS
quasars. In section \ref{section:nebulae}, we present the discovery of large nebulae in the quasar fields.
In section \ref{section:discussion}, we discuss the implication of our findings, search for correlations
between the nebulae and quasar properties, and compare with circum-quasar nebulae at higher redshift. Finally, we summarize
our results and make concluding remarks in Section \ref{section:conclusions}.

Throughout, we adopt a flat $\Lambda$ cosmology with $\Omega_{\rm m}=0.3$,
$\Omega_\Lambda=0.7$, and a Hubble constant of $H_0=70$ \kms. Unless otherwise
stated, all magnitudes are given in the AB system \citep[][]{Oke:1983}, all wavelengths
are given in vacuum, and all distances are proper.

\section{Observations, data reduction, and processing}
\label{section:observations}
Both the CUBS and MUSEQuBES surveys were designed to study the CGM and IGM
around galaxies at $z\approx 0.1-1.4$ using deep and highly complete galaxy redshift
surveys with MUSE in the fields of UV-bright background quasars with high-quality UV
absorption spectra from the Cosmic Origins Spectrograph \citep{Green:2012} aboard the
{\it Hubble Space Telescope}. The broad spectral coverage ($4700-9350$ \AA), wide
field-of-view ($1'\times1'$), and high throughput (peaking at 35\% at $7000$ \AA)
of MUSE \cite[][]{Bacon:2010} collectively enable searches for low surface brightness
nebulae in the environments of the CUBS and MUSEQuBES background quasars
at $z=0.4-1.4$ \citep[][]{Johnson:2018, Johnson:2022, Chen:2023, Liu:2024} {at the same time}.
The MUSEQuBES MUSE observations (PI: J. Schaye) were conducted in the MUSE wide-field mode
under natural seeing conditions characterized by a  full width at half maximum of 
${\rm FWHM}=0.5-1.0''$ with two to ten hours of integration.
The CUBS MUSE observations (PI: H.-W. Chen) have integration times ranging
from 1.4 to 3.4 hours and were conducted in wide-field mode with ground-layer adaptive
optics \citep[][]{Kolb:2016, Madec:2018}. As a result, the CUBS datacubes exhibit improved
image quality of ${\rm FWHM}=0.5-0.7''$.

\begin{table*}
\caption{Summary of quasar properties and MUSE observations, sorted by quasar redshift.}
\label{table:quasars}
\centering
\begin{tabular}{lrrcccccrcrc}
\hline
\hline
& & \multicolumn{8}{c}{quasar properties} & \multicolumn{2}{c}{MUSE data} \\
  \cmidrule(lr){2-10}   \cmidrule(lr){11-12}
&  \multicolumn{1}{c}{R.A.$^a$}            &  \multicolumn{1}{c}{Decl.$^b$}              &            	      &    &  {$\log \lambda L_{\rm \lambda}^e$}          & $\log L_{\rm bol}^f$  & \multicolumn{1}{c}{$\log M_{\rm BH}^g$} &  & {$W_{\rm r}([{\rm O\,II]})^i$} & \multicolumn{1}{c}{$t_{\rm exp}^j$} & seeing$^k$ \\
\multicolumn{1}{c}{quasar name}          &  \multicolumn{1}{c}{(deg)}       &  \multicolumn{1}{c}{(deg)}          &  $z$$^c$	      & line$^d$   & {$/{\rm erg\,s^{-1}}$}     & $/{\rm erg\,s^{-1}}$  & $/{\rm M}_{\odot}$   & $\log RL^h$ & {(\AA)}   & \multicolumn{1}{c}{(hr)} & (arcsec)  \\
\hline
HE\,0435$-$5304      &  $ 69.2118  $   &  $-52.9800 $   &  $0.4279$        & [O\,II]  & {$44.9$} & $45.8$   & $8.4 $  &  $<1.0$  & {$-0.78\pm0.04$} & $ 2.0$ & $1.0$ \\
HE\,0153$-$4520      &  $ 28.8050  $   &  $-45.1032 $   &  $0.4532$        & H$\beta$ & {$45.5$} & $46.5$   & $8.9 $  &  $<0.9$  & {$-0.01\pm0.01$} & $ 2.0$ & $0.7$ \\
HE\,0226$-$4110      &  $ 37.0635  $   &  $-40.9541 $   &  $0.4936$        & [O\,II]  & {$46.0$} & $47.0$   & $9.0 $  &  $<0.1$  & {$-0.17\pm0.01$} & $ 8.0$ & $0.5$ \\
PKS\,0405$-$123      &  $ 61.9518  $   &  $-12.1935 $   &  $0.5731$        & [O\,II]  & {$46.4$} & $47.3$   & $9.4 $  &  $3.1$   & {$-0.36\pm0.01$} & $ 9.8$ & $1.0$ \\
HE\,0238$-$1904      &  $ 40.1357  $   &  $-18.8642 $   &  $0.6282$        & [O\,II]  & {$46.2$} & $47.2$   & $9.8 $  &  $<0.3$  & {$-0.34\pm0.01$} & $10.0$ & $0.7$ \\
3C\,57               &  $ 30.4882  $   &  $-11.5426 $   &  $0.6718$        & [O\,II]  & {$45.8$} & $46.8$   & $8.9 $  &  $3.7$   & {$-0.55\pm0.02$} & $ 2.0$ & $0.7$ \\
PKS\,0552$-$640      &  $ 88.1020  $   &  $-64.0363 $   &  $0.6824$        & [O\,II]  & {$46.4$} & $47.4$   & $9.7 $  &  $2.3$   & {$-0.32\pm0.02$} & $ 2.0$ & $0.7$ \\
J0110$-$1648         &  $ 17.6479  $   &  $-16.8077 $   &  $0.7822$        & [O\,II]  & {$46.2$} & $46.9$   & $9.1 $  &  $2.4$   & {$-0.23\pm0.02$} & $ 1.4$ & $0.6$ \\
J0454$-$6116         &  $ 73.5664  $   &  $-61.2740 $   &  $0.7864$        & [O\,II]  & {$46.2$} & $46.9$   & $9.4 $  &  $<0.9$  & {$-0.84\pm0.02$} & $ 1.4$ & $0.7$ \\
J2135$-$5316         &  $ 323.9716 $   &  $-53.2821 $   &  $0.8123$        & [O\,II]  & {$46.6$} & $47.3$   & $9.4 $  &  $<0.6$  & {$-0.22\pm0.01$} & $ 1.9$ & $0.5$ \\
J0119$-$2010         &  $ 19.9837  $   &  $-20.1729 $   &  $0.8160$        & [O\,II]  & {$46.4$} & $47.1$   & $9.3 $  &  $<0.9$  & {$-0.27\pm0.01$} & $ 1.4$ & $0.5$ \\
HE\,0246$-$4101      &  $ 42.0261  $   &  $-40.8093 $   &  $0.8840$        & [O\,II]  & {$46.7$} & $47.4$   & $9.7 $  &  $<0.7$  & {$-0.22\pm0.03$} & $ 2.1$ & $0.7$ \\
J0028$-$3305         &  $ 7.1266   $   &  $-33.0970 $   &  $0.8915$        & [O\,II]  & {$46.3$} & $47.0$   & $9.4 $  &  $<1.0$  & {$-0.13\pm0.01$} & $ 1.4$ & $0.5$ \\
HE\,0419$-$5657      &  $ 65.2246  $   &  $-56.8455 $   &  $0.9481$        & [O\,II]  & {$46.1$} & $46.8$   & $9.2 $  &  $<1.2$  & {$-0.27\pm0.05$} & $ 1.4$ & $0.6$ \\
Q0107$-$025          &  $ 17.5677  $   &  $-2.3141  $   &  $0.9545$        & [O\,II]  & {$46.2$} & $46.9$   & $9.4 $  &  $<1.1$  & {$-0.67\pm0.10$} & $ 2.0$ & $0.9$ \\
Q0107$-$0235         &  $ 17.5548  $   &  $-2.3313  $   &  $0.9574$        & [O\,II]  & {$46.1$} & $46.8$   & $9.6 $  &  $3.2$   & {$-1.36\pm0.05$} & $ 2.0$ & $0.8$ \\
PKS\,2242$-$498      &  $ 341.2508 $   &  $-49.5301 $   &  $1.0011$        & [O\,II]  & {$46.3$} & $47.0$   & $9.6 $  &  $3.4$   & {$-0.83\pm0.03$} & $ 1.4$ & $0.6$ \\
PKS\,0355$-$483      &  $ 59.3413  $   &  $-48.2042 $   &  $1.0128$        & [O\,II]  & {$46.3$} & $47.0$   & $9.6 $  &  $2.9$   & {$-0.84\pm0.02$} & $ 2.6$ & $0.5$ \\
HE\,0112$-$4145      &  $ 18.5921  $   &  $-41.4964 $   &  $1.0238$        & [O\,II]  & {$46.2$} & $46.9$   & $9.5 $  &  $1.2$   & {$-2.04\pm0.05$} & $ 1.4$ & $0.7$ \\
HE\,0439$-$5254      &  $ 70.0502  $   &  $-52.8048 $   &  $1.0530$        & Mg\,II   & {$46.3$} & $47.0$   & $9.3 $  &  $<1.1$  & {$-0.03\pm0.04$} & $ 2.5$ & $0.7$ \\
HE\,2305$-$5315      &  $ 347.1574 $   &  $-52.9802 $   &  $1.0733$        & H$\beta$ & {$46.5$} & $47.2$   & $9.5 $  &  $<1.0$  & {$-0.03\pm0.02$} & $ 1.9$ & $0.6$ \\
HE\,1003$+$0149      &  $ 151.3968 $   &  $+1.5793  $   &  $1.0807$        & [O\,II]  & {$46.3$} & $47.0$   & $9.3 $  &  $1.5$   & {$-0.62\pm0.03$} & $ 2.0$ & $0.9$ \\
HE\,0331$-$4112      &  $ 53.2794  $   &  $-41.0336 $   &  $1.1153$        & [O\,III] & {$46.8$} & $47.5$   & $9.9 $  &  $<0.8$  & {$-0.11\pm0.03$} & $ 2.6$ & $0.6$ \\
TXS\,0206$-$048      &  $ 32.3781  $   &  $-4.6406  $   &  $1.1317$        & [O\,II]  & {$46.4$} & $47.2$   & $9.8 $  &  $3.0$   & {$-1.65\pm0.03$} & $ 8.0$ & $0.6$ \\
Q1354$+$048          &  $ 209.3594 $   &  $+4.5948  $   &  $1.2335$        & [O\,II]  & {$46.5$} & $47.2$   & $9.5 $  &  $<1.1$  & {$-0.67\pm0.04$} & $ 2.0$ & $0.5$ \\
J0154$-$0712         &  $ 28.7278  $   &  $-7.2061  $   &  $1.2957$        & [O\,II]  & {$46.8$} & $47.5$   & $9.5 $  &  $<0.9$  & {$-0.11\pm0.01$} & $ 3.4$ & $0.5$ \\
Q1435$-$0134         &  $ 219.4511 $   &  $-1.7863  $   &  $1.3117$        & [O\,II]  & {$47.1$} & $47.8$   & $10.0$  &  $1.8$   & {$-0.41\pm0.02$} & $ 6.1$ & $0.6$ \\
PG\,1522$+$101       &  $ 231.1021 $   &  $+9.9747  $   &  $1.3302$        & [O\,II]  & {$47.0$} & $47.7$   & $10.1$  &  $<0.7$  & {$-0.18\pm0.02$} & $ 2.0$ & $0.5$ \\
HE\,2336$-$5540      &  $ 354.8050 $   &  $-55.3974 $   &  $1.3531$        & [O\,II]  & {$47.1$} & $47.8$   & $10.0$  &  $2.5$   & {$-0.23\pm0.01$} & $ 2.6$ & $0.6$ \\
PKS\,0232$-$04       &  $ 38.7805  $   &  $-4.0348  $   &  $1.4450$        & [O\,II]  & {$46.7$} & $47.4$   & $9.7 $  &  $4.0$   & {$-0.70\pm0.12$} & $10.0$ & $0.7$ \\
\hline
\multicolumn{12}{l}{$^a$Right ascension of the quasar in J2000 coordinates from the Gaia survey Data Release 3 \citep[][]{Gaia-Collaboration:2023}.}\\
\multicolumn{12}{l}{$^b$Declination of the quasar in J2000 coordinates from the Gaia survey Data Release 3 \citep[][]{Gaia-Collaboration:2023}.}\\
\multicolumn{12}{l}{$^c$Quasar systemic redshift measured as described in Section \ref{section:quasars}.}\\
\multicolumn{12}{l}{$^d$Emission line used to measure the quasar systemic redshift. {Systematic redshift uncertainties correspond to $\approx 20, 50,$ and}}\\
\multicolumn{12}{l}{\, {$130$ \kms\ for [O\,II], [O\,III], and H$\beta$/Mg\,II-based redshifts, respectively.}} \\
\multicolumn{12}{l}{$^e$Monochromatic luminosity measured at rest-frame $\lambda=5100$ and $3000$ \AA\ for quasars at $z<0.7$ and $z>0.7$, respectively.}\\
\multicolumn{12}{l}{\,  {Systematic uncertainties in the monochromatic luminosities are dominated by $\approx 10\%$ flux calibration errors.}} \\
\multicolumn{12}{l}{$^f$Bolometric luminosity of the quasar estimated as described in Section \ref{section:quasars}. {Systematic uncertainty in bolometric luminosities are}}\\
\multicolumn{12}{l}{\,  {dominated by population scatter in the bolometric corrections of $\approx 0.1$ dex \citep[][]{Richards:2006}.}} \\
\multicolumn{12}{l}{$^g$Supermassive black hole mass estimated with a single-epoch virial theorem approach described in Section \ref{section:quasars}. {Systematic}}\\
\multicolumn{12}{l}{\, {uncertainties in these black hole mass estimates are $\approx 0.4$ dex \citep[][]{Shen:2023}.}}\\
\multicolumn{12}{l}{$^h$Radio-loudness based on flux at rest-frame wavelengths of 6 cm and 2500 \AA. {Systematic uncertainties in $\log RL$ are $\approx 0.1$ dex.}}\\
\multicolumn{12}{l}{$^i${Rest-frame [O\,II] equivalent widths measured in the MUSE quasar spectra within the seeing disk corresponding to $\lesssim5$ kpc.}}\\
\multicolumn{12}{l}{$^j$Total exposure time of the MUSE observations.}\\
\multicolumn{12}{l}{$^k$FWHM seeing measured in the MUSE white-light image of the quasar.}\\
\end{tabular}
\end{table*}

\begin{figure}
\includegraphics[width=\columnwidth]{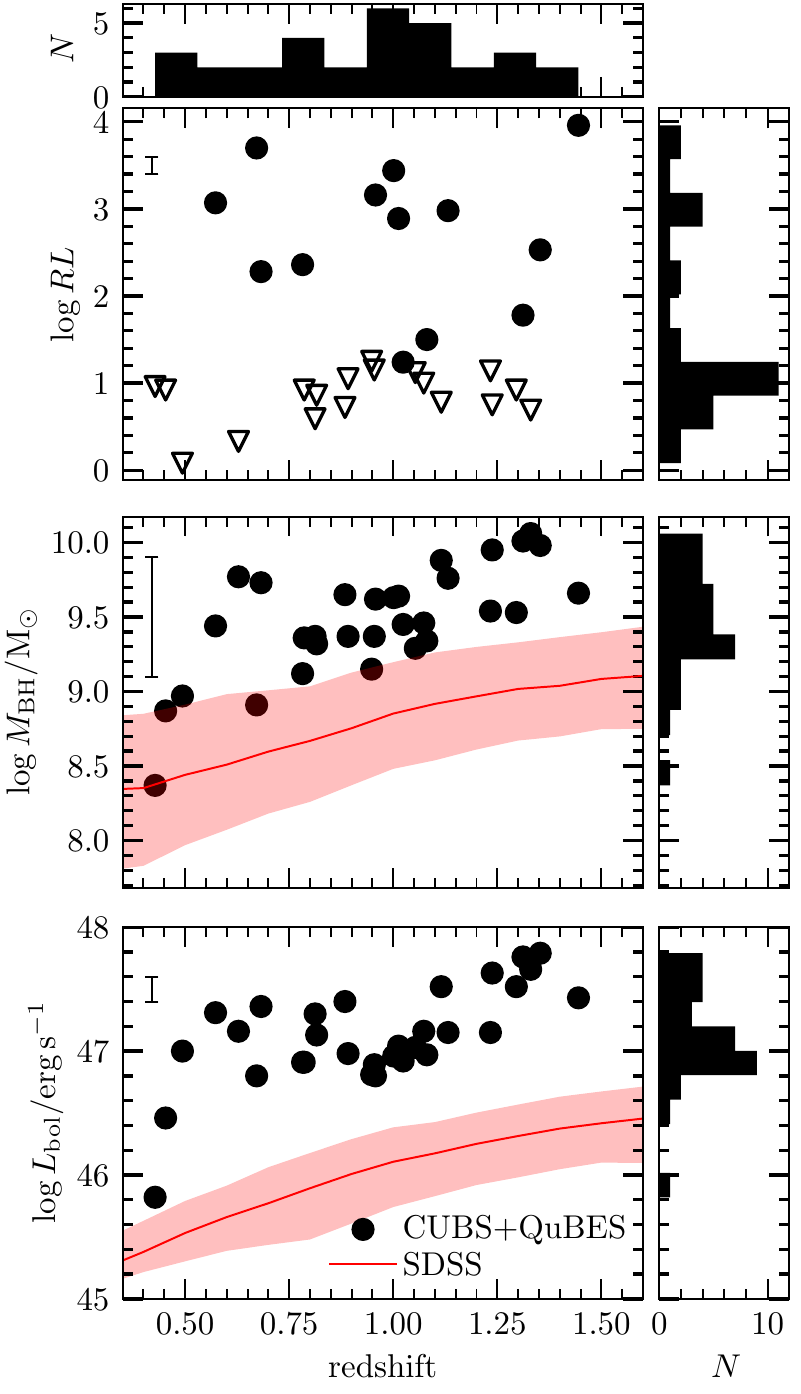}
\caption{Summary of the properties of the CUBS and MUSEQuBES quasars.
The main panels show bolometric luminosity ({\it bottom}), black hole mass ({\it middle}), and radio-loudness ({\it top}) 
of the CUBS and MUSEQuBES quasars versus redshift as black points. Quasars not detected in radio data
are shown as open symbols marking the $3\sigma$ upper limits on their radio-loudness. {Uncertainties are dominated by systematics and visualized as error bars in top left of each panel.}
The median bolometric luminosity and black hole mass for SDSS quasars \citep[][]{Shen:2011} is shown as a function of redshift
with a red line in the bottom and middle panels, with a faded red band marking 68\% population scatter
at each redshift. The small panels display histograms of the redshifts ({\it top}), bolometric luminosities ({\it bottom right}),
black hole masses ({\it middle right}), and radio-loudnesses ({\it top right}). By selection,
the CUBS and MUSEQuBES quasars are more luminous and have higher
inferred black hole masses than typical SDSS quasars at the same redshifts. The sample is evenly
split between radio-quiet and loud systems.}
\label{figure:quasars}
\end{figure}

We reduced the MUSEQuBES data with the MUSE Guaranteed Time Observations (GTO) team
pipeline as described in \cite{Bacon:2017}, which includes standard reduction steps in the MUSE ESO
pipeline \citep[][]{Weilbacher:2020} but with post-processing that improves image quality via self-calibration
and  principle component-based sky subtraction \citep[][]{Soto:2016}. We also
reduced the data using the CubEx post-processing pipeline described in \cite{Borisova:2016} and \cite{Cantalupo:2019}.
For MUSEQuBES fields, we performed the analysis in this paper on both
the GTO and CubEx reduction products and found consistent results in all cases.

We reduced the CUBS datacubes with the ESO pipeline both with
and without CubEx post-processing as described in \cite{Chen:2020} and \cite{Chen:2023}.
For all of the CUBS and MUSEQuBES fields, we also retrieved the reduced, coadded MUSE-DEEP
datacubes from the ESO archive. Compared to the GTO and CubEx data products,
MUSE-DEEP reductions exhibit more artifacts in continuum images and larger spectral residuals near sky lines.
However, for the narrow-band, continuum-subtracted {datacubes and} surface brightness maps used in this work,
we obtained similar results with the ESO-DEEP products except when near a night sky emission line, in which case the CubEx
and GTO reductions exhibit significantly improved residuals. After reducing the data, we converted the
air wavelengths delivered by the MUSE pipelines to vacuum for convenience. The total MUSE exposure time available and seeing
FWHM measured in the final stacked datacubes are listed for each quasar field in Table \ref{table:quasars}.

{
The ESO, GTO, and CubEx pipelines all share common wavelength and flux calibration approaches. MUSE wavelength calibration 
frames are taken with three built-in lamps
 (HgCd, Ne, and Xe) by ESO Science Operations staff in the morning, after the science exposures are acquired.
The observed locations of the arc lines on the detectors are fit with outlier-resistant 2D polynomials
to simultaneously solve for the dispersion, tilt, and curvature of the spectral lines in the arc frames.
Comparisons of arc lamp spectra taken over the course of a night and observations of telluric absorption features
 in MUSE spectra of bright stars indicate that systematic uncertainties in the wavelength calibration are typically
less than $1$ \kms\ \citep[][]{Kamann:2018, Weilbacher:2020}. 
}

{
MUSE flux calibrations are taken with spectrophotometric standards nightly.
Systematic uncertainties in the spectrophotometry as a function of position in the datacube and wavelength are typically less than 5\%
for data take under photometric conditions \citep[][]{Weilbacher:2020}, though others have reported systematic errors in the flux calibration of up to 25\% \citep[e.g.][]{Peroux:2019}. To quantify the level of systematic errors in the CUBS and MUSEQuBES flux calibration, we identified bright stars and galaxies in each MUSE datacube to serve as references and calculated synthetic photometry in the ACS$+$F814W and Dark Energy Camera $i$-bands. We then compared these synthetic magnitudes with those measured in archival HST ACS$+$F814W images of the MUSEQuBES fields and public Dark Energy Survey Data Release 2 \citep[DES;][]{Abbott:2021} photometry for the CUBS fields. Based on comparisons of the synthetic MUSE photometry and reference photometry from HST and DES, we estimate a $1\sigma$ uncertainty of 10\%.
}

{
One of the primary differences in the ESO, GTO, and CubEx pipelines is their approach to illumination corrections.
The ESO pipeline performs illumination corrections using a combination of flats taken with the calibration unit \citep[][]{Kelz:2012} and
twilight flats. However, subtle differences in the true illumination pattern during the science observations and those taken with the
calibration unit and twilight sky correlate with IFU slice, resulting in artificial grid patterns in
narrow-band images and emission line maps \citep[see discussion in ][]{Bacon:2015, Borisova:2016}.
To improve the illumination correction and remove these artificial patterns from MUSE data, both the GTO and CubEx
pipelines perform ``self-calibration'' by producing illumination corrections based on the night sky continuum and emission lines
observed in science frames after masking continuum sources. These self-calibration steps effectively remove the grid pattern present in ESO cubes
and reduce residuals in synthetic images formed from the data cubes by a factor of approximately three \citep{Lofthouse:2020}.
}

{
The other major differences in the ESO, GTO, and CubEx pipeline products results from their night-sky
subtraction approaches. All three pipelines begin by masking continuum sources to identify spaxels where
the detected flux is dominated by sky emission, but their modeling and subtraction of the night sky emission
differ significantly. The ESO sky subtraction pipeline adopts a model of the night sky-line emission based on a fixed set of 5100
cataloged emission lines \citep[][]{Cosby:2006} in the MUSE wavelength range. These are categorized into 52 groups,
with lines in each group sharing the same excited state. The night-sky line emission model fixes the relative intensity of lines
in each of these groups and fits the observed global sky spectrum, allowing the total intensity of each group to vary and convolving
with the line-spread function (LSF). After this, the best-fitting global sky model is subtracted and the residuals in source-free spaxels are used to estimate
the continuous component of the night sky emission for each slice \citep[][]{Streicher:2011}.}

{To reduce night sky residuals present after the ESO sky subtraction,
the GTO pipeline employs post-processed
sky subtraction with principal component analysis using the Zurich Atmosphere Purge package \citep[ZAP;][]{Soto:2016}.
On the other hand, the CubEx pipeline operates with post-processing on ESO data products with the ESO sky subtraction module turned off.
It then uses the CubeSharp \citep[][]{Borisova:2016, Cantalupo:2019}
routine to produce an empirical model of the LSF on a line-by-line basis (or groups of lines if they are close in wavelength).
One key advantage of the CubEx sky subtraction approach is that the LSF can be finely adjusted to improve sky subtraction while also conserving flux.
Both the GTO and CubEx pipelines improve sky line residuals significantly, resulting in a factor of approximately three reduction in the
standard deviation of flux in source-free spaxels \citep[][]{Lofthouse:2020}.
}

To study faint line emission near the bright quasars, we performed quasar light subtraction
as described in \cite{Johnson:2018} and \cite{Helton:2021}, which effectively removes the spatially unresolved
continuum, broad-line emission, and narrow-line emission from the nucleus.  Finally,
to search for line-emitting nebulae, we produced continuum-subtracted emission-line maps around the expected
wavelength of the [O\,II] doublet at the redshift of each quasar. To do so, we first identified suitable continuum
regions on the blue and red sides of the potential emission line. The continuum regions are typically $3000{-}4000$
\kms\ in width but tailored for each quasar based on proximity to night-sky emission lines. We then fit a low-order
polynomial to the continuum regions in each spaxel of the datacube and subtracted the best-fit model to produce
continuum-subtracted datacubes using the MUSE Python Data Analysis Framework \citep[MPDAF][]{Bacon:2016}.

\section{Quasar Properties}
\label{section:quasars}
To characterize the properties of the CUBS and MUSEQuBES quasars, we extracted their optical MUSE 
spectra with MPDAF \citep[][]{Bacon:2016, Piqueras:2017} prior to quasar and continuum subtraction with apertures
chosen to include $>95\%$ of the quasar light. We measured precise, narrow-line redshifts for the quasars
using the [O\,II] $\lambda\lambda3727,3729$ doublet whenever possible, assuming an effective centroid of
$\lambda 3728.6$ expected for the $0.8{:}1$ to $0.9{:}1$ doublet ratio typical of quasar narrow-line
emission \citep[][]{Hewett:2010}. {At the same time, we also measured the rest-frame equivalent width of the [O\,II]
doublet in the quasar spectra extracted within a seeing disk, $W_{\rm r}([{\rm O\,II}])$. The seeing disks correspond to
$\lesssim 5$ kpc from the quasars, ensuring that the [O\,II] equivalent widths capture emission from the narrow-line region rather than more extended nebulae}. Quasar redshifts based on [O\,II] emission are characterized by
typical uncertainties of $20-30$ \kms\ \citep[][]{Boroson:2005, Hewett:2010}. Four quasars do not
exhibit detectable [O\,II] despite the high S/N available in the MUSE spectra, necessitating the use of other
emission line observed in the near-IR by \cite{Sulentic:2004}. For HE\,0331$-$4112, we estimated the systemic redshift using the [O\,III] $\lambda 5008$
emission-line peak. Quasar redshifts measured with the peak of the [O\,III] line are characterized by typical systematic
uncertainties of $\approx 50$ \kms\ when compared to stellar absorption-based redshifts
\cite[see discussion in][]{Shen:2016}. For the remaining three quasars, narrow emission lines are not detectable,
necessitating the use of broad H$\beta$ (HE\,0153$-$4520 and HE\,2305$-$5315) or Mg\,II (HE\,0439-5254).
We estimated the systemic redshifts of these three quasars {by cross-correlating the observed quasar spectra with the template
from \cite{Hewett:2010} using methods described in that work. As discussed in \cite{Hewett:2010}, the
resulting broad H$\beta$ and Mg\,II emission redshifts exhibit uncertainties of $\approx 130$ \kms}
when compared to more precise narrow-line redshifts.
The quasar redshifts and the emission lines driving them are listed in Table \ref{table:quasars}.

To estimate the luminosities and black hole masses of the quasars, we used PyQSOFit \citep[][]{Guo:2018, Guo:2019}
to fit the quasar continuum with a power-law and Fe\,II templates \citep[][]{Boroson:1992, Vestergaard:2001}
and multiple Gaussians for broad and narrow emission lines.
To enable estimates of the bolometric luminosities and black hole masses of the quasars, we measured the
monochromatic luminosity at $5100$ \AA\ and line-width of broad H$\beta$ for systems at $z<0.7$,
and the monochromatic luminosity at $3000$ \AA\ and line-width of broad Mg\,II for those at $z>0.7$.
For four CUBS quasars (HE\,0331$-$4112, J2135$-$5316, PKS\,2242$-$498, and HE\,2305$-$5315), the Mg\,II emission-line falls
near or in the MUSE Sodium laser gap necessitated by the adaptive optics system. For these, we obtained supplementary
spectra with the Magellan Echellete (MagE) spectrograph \citep[][]{Marshall:2008} to enable Mg\,II line-width measurements
(for details of the MagE observations and reductions, see \cite{Li:2024}). We then combined the monochromatic luminosities and line-widths to infer supermassive black hole masses
 ($M_{\rm BH}$) using the single-epoch virial theorem relations described in \cite{Shen:2011}.
 Typical uncertainties in single-epoch black hole mass estimates are $0.4$ dex, though uncertainties may be higher
for the CUBS and MUSEQuBES quasars because they are higher in luminosity than typical reverberation mapping quasar samples
used to calibrate the relations \citep[see discussion in ][]{Shen:2023}.
 
 To enable comparisons of the luminosities
 of the quasars across redshifts, we converted the monochromatic luminosity measurements to bolometric luminosity estimates ($L_{\rm bol}$)
 using bolometric corrections {from \cite{Richards:2006} of $10.3\times$ and $5.6\times$ for monochromatic luminosities at 
 $5100$ \AA\ and $3000$ \AA, respectively.
Uncertainties in the bolometric luminosities are dominated by $\approx 0.1$ dex intrinsic scatter in bolometric corrections  \citep[][]{Richards:2006}.}
Finally, we estimated the radio-loudness (RL), defined as the flux density ratio
at rest-frame 6 cm vs. 2500 \AA, of each quasar using radio observations
from the Rapid Australian Square Kilometre Array Pathfinder (ASKAP) Continuum Survey (RACS) Data Release 1 \citep[][]{Hale:2021}
as described in \cite{Li:2024}. {In the case of non-detections, we calculated $3\sigma$ upper limits on the radio flux
and converted these to upper limits on radio-loudness.}
{The RACS radio observations and MUSE rest-NUV observations are separated by a time period of several years,
resulting in non-negligible systematic uncertainty of 0.1 dex due to potential NUV variability on these timescales \citep[][]{Welsh:2011},
though we caution that rare but significant radio variability could result in significantly larger errors in the radio-loudness \citep[][]{Nyland:2020}.}
The {monochromatic luminosities}, bolometric luminosities, supermassive black hole mass estimates, radio-loudness measurements,
{and [O\,II] rest-frame equivalent widths}
for each quasar are reported in Table \ref{table:quasars}.

To contextualize the properties of the CUBS and MUSEQuBES quasars, we plot their
luminosities, black hole masses, and radio-loudnesses vs redshift and compare with quasars
in the SDSS sample from \cite{Shen:2011} in Figure \ref{figure:quasars}. The CUBS and MUSEQuBES
quasars were selected to be sufficiently UV bright for high-quality absorption spectroscopy with COS.
As a result, the CUBS and MUSEQuBES quasars are among the most luminous quasars in the $z<1.4$
Universe, with luminosities typically $\approx1.5$ dex higher than the median SDSS quasar at the same redshift.
Similarly, the inferred SMBH masses of the CUBS and MUSEQuBES quasars are $\approx 1$ dex
above the SDSS median at similar redshifts \citep{Shen:2011}. The 30 quasars are approximately evenly split between
radio-loud and radio-quiet with a division between the two classes of $\log RL \approx 1$.

\section{Discovery of optically emitting circumgalactic nebulae}
\label{section:nebulae}

\begin{figure*}
\centering
\includegraphics[width=0.95\textwidth]{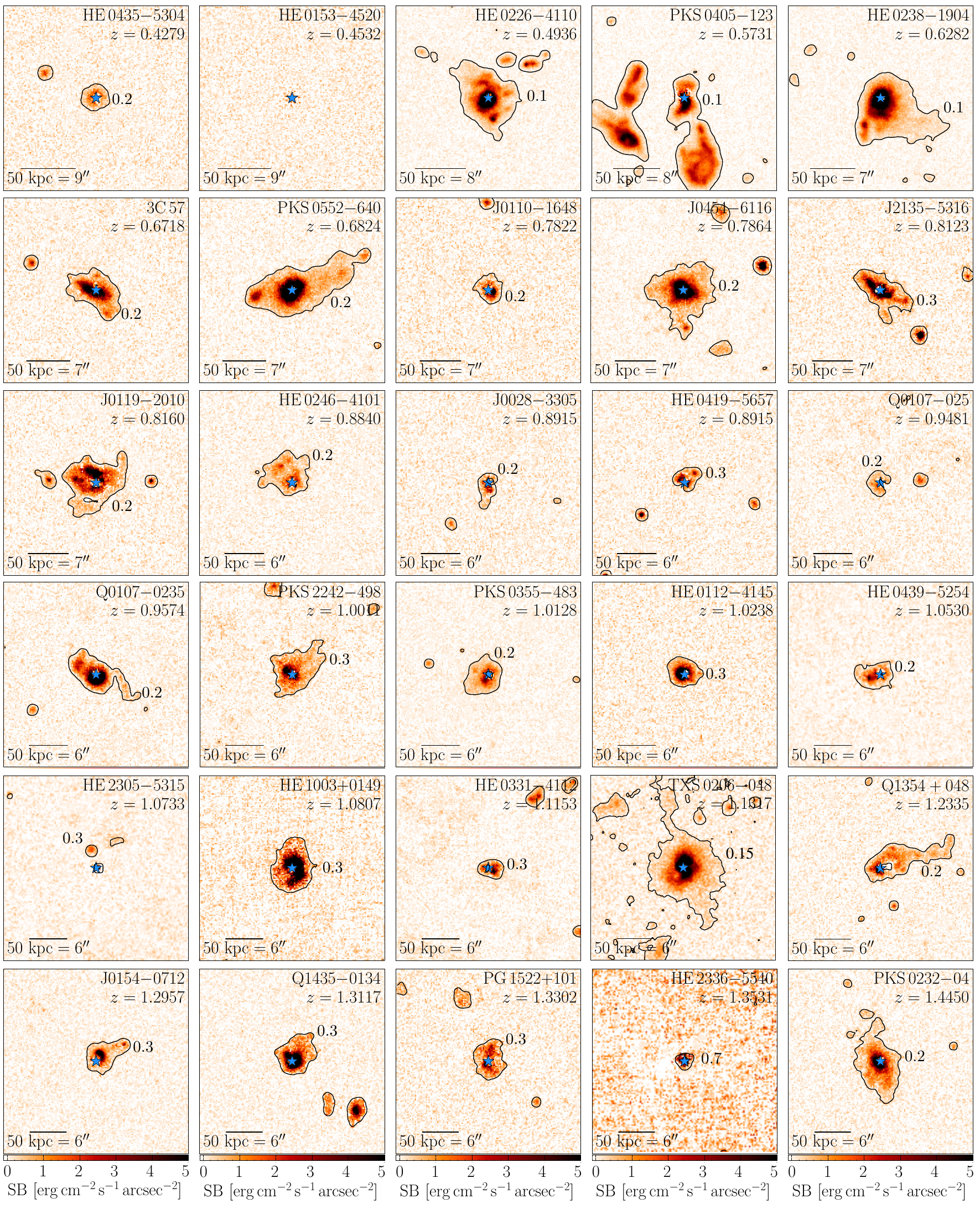}\vspace{-0.3cm}
\caption{[O\,II] surface brightness maps for each quasar field ordered by redshift. The surface brightness
maps are displayed after quasar light subtraction, continuum subtraction, and 3D extraction, as described in the text. Each panel is 30''$\times$30'' on a side,
and a blue star at the center of each image marks the quasar's position. A scale bar corresponding to $50$ proper kpc at the redshift of the quasar is shown in the bottom left of each panel. The names and redshifts of the quasars are shown in the top right of each panel. The contour line in each panel corresponds to the {$3\sigma$} detection limit averaged over one square arcsecond based on the variance in each image and is labeled in units of $10^{-17}\ {\rm erg\,s^{-1}\,cm^{-2}\,arcsec^{-2}}$. {The contours are included to visualize the extent of the nebulae and are distinct from the $5\sigma$ significance requirement used to define a detection.}}

\label{figure:SB_all}
\end{figure*}

\begin{table*}
\caption{Summary of the [O\,II] emitting nebulae around the CUBS and MUSEQuBES quasars, ordered by redshift.}
\label{table:nebulae}
\centering
\begin{tabular}{lcccrrrcc}
\hline
\hline
                    &          &        & FWHM$^c$     & size$^d$   & area$^e$        &        & [O\,II] luminosity$^g$       & ${\rm SB}_{10-30}(1 + z)^{4\, h}$      \\
quasar              & $z$$^a$ &   flag$^b$ & (arcsec) & (kpc)  & (kpc$^2$)   &   SNR$^f$  & ($\rm erg\,s^{-1}$)      & ($\rm erg\,cm^{-2}\,s^{-1}\,arcsec^{-2}\,$)  \\
\hline
HE\,0435$-$5304     & 0.4279   &   $=$  &  2.6     &  55    &  1170      &   35.3 &   $ 9.6{\times}10^{41} $   & $5.6{\times}10^{-17}$    \\
HE\,0153$-$4520     & 0.4532   &   $<$  &  0.6     &   7    &    35      &    2.4 &   $ 1.9{\times}10^{40} $   & $1.4{\times}10^{-18}$   \\
HE\,0226$-$4110     & 0.4936   &   $=$  &  3.9     &  84    &  2740      &  156.3 &   $ 1.1{\times}10^{42} $   & $3.9{\times}10^{-17}$    \\
PKS\,0405$-$123$^i$     & 0.5731   &   $=$  &  9.1     & 129    &  8180      &  134.8 &   $ 2.7{\times}10^{42} $   & $2.2{\times}10^{-17}$   \\
HE\,0238$-$1904$^j$     & 0.6282   &   $=$  &  4.9     & 103    &  4820      &  187.9 &   $ 2.3{\times}10^{42} $   & $9.0{\times}10^{-17}$       \\
3C\,57              & 0.6718   &   $=$  &  3.6     &  71    &  1930      &   71.3 &   $ 1.2{\times}10^{42} $   & $5.6{\times}10^{-17}$    \\
PKS\,0552$-$640     & 0.6824   &   $=$  &  6.2     & 153    &  5180      &  119.6 &   $ 3.4{\times}10^{42} $   & $1.1{\times}10^{-16}$    \\
J0110$-$1648        & 0.7822   &   $=$  &  1.8     &  29    &   530      &   20.3 &   $ 4.1{\times}10^{41} $   & $1.3{\times}10^{-17}$    \\
J0454$-$6116$^k$        & 0.7864   &   $=$  &  3.5     & 102    &  4350      &  234.0 &   $ 4.8{\times}10^{42} $   & $1.3{\times}10^{-16}$      \\
J2135$-$5316$^k$        & 0.8123   &   $=$  &  3.8     &  83    &  2250      &   47.9 &   $ 2.1{\times}10^{42} $   & $9.4{\times}10^{-17}$      \\
J0119$-$2010        & 0.8160   &   $=$  &  4.2     &  96    &  3090      &   60.1 &   $ 2.5{\times}10^{42} $   & $1.4{\times}10^{-16}$    \\
HE\,0246$-$4101     & 0.8840   &   $=$  &  3.8     &  74    &  2280      &   44.3 &   $ 1.0{\times}10^{42} $   & $4.6{\times}10^{-17}$   \\
J0028$-$3305        & 0.8915   &   $=$  &  2.1     &  42    &   500      &   23.2 &   $ 2.9{\times}10^{41} $   & $1.1{\times}10^{-17}$    \\
HE\,0419$-$5657     & 0.9481   &   $=$  &  2.1     &  35    &   620      &   23.2 &   $ 5.6{\times}10^{41} $   & $2.5{\times}10^{-17}$    \\
Q0107$-$025         & 0.9545   &   $=$  &  1.9     &  28    &   410      &   13.9 &   $ 2.0{\times}10^{41} $   & $1.4{\times}10^{-17}$    \\
Q0107$-$0235   & 0.9574   &   $=$  &  3.1     &  90    &  2130      &  144.8 &   $ 3.7{\times}10^{42} $   & $1.3{\times}10^{-16}$    \\
PKS\,2242$-$498     & 1.0011   &   $=$  &  3.4     &  71    &  2260      &   54.6 &   $ 2.3{\times}10^{42} $   & $1.2{\times}10^{-16}$    \\
PKS\,0355$-$483     & 1.0128   &   $=$  &  2.8     &  50    &  1310      &   70.5 &   $ 1.0{\times}10^{42} $   & $6.2{\times}10^{-17}$    \\
HE\,0112$-$4145     & 1.0238   &   $=$  &  1.8     &  38    &   890      &   70.8 &   $ 2.7{\times}10^{42} $   & $7.8{\times}10^{-17}$    \\
HE\,0439$-$5254     & 1.0530   &   $=$  &  2.5     &  47    &   970      &   34.3 &   $ 9.1{\times}10^{41} $   & $5.6{\times}10^{-17}$    \\
HE\,2305$-$5315     & 1.0733   &   $<$  &  0.6     &  15    &    100      &    3.9 &   $ 9.6{\times}10^{40} $   & $1.1{\times}10^{-17}$    \\
HE\,1003$+$0149     & 1.0807   &   $=$  &  2.8     &  53    &  1660      &   94.8 &   $ 7.0{\times}10^{42} $   & $3.9{\times}10^{-16}$    \\
HE\,0331$-$4112     & 1.1153   &   $=$  &  1.8     &  32    &   520      &   25.9 &   $ 6.9{\times}10^{41} $   & $3.1{\times}10^{-17}$    \\
TXS\,0206$-$048$^l$     & 1.1317   &   $=$  &  5.0     & 200    & 10800      &  257.1 &   $ 9.7{\times}10^{42} $   & $3.0{\times}10^{-16}$   \\
Q1354$+$048         & 1.2335   &   $=$  &  6.4     & 126    &  2860      &   66.5 &   $ 2.9{\times}10^{42} $   & $9.9{\times}10^{-17}$    \\
J0154$-$0712        & 1.2957   &   $=$  &  2.8     &  63    &  1340      &   41.9 &   $ 2.2{\times}10^{42} $   & $1.2{\times}10^{-16}$    \\
Q1435$-$0134        & 1.3117   &   $=$  &  2.7     &  63    &  2010      &   76.0 &   $ 7.3{\times}10^{42} $   & $2.6{\times}10^{-16}$    \\
PG\,1522$+$101      & 1.3302   &   $=$  &  2.9     &  50    &  1220      &   28.3 &   $ 2.5{\times}10^{42} $   & $1.6{\times}10^{-16}$    \\
HE\,2336$-$5540     & 1.3531   &   $<$  &  0.4     &   7    &    30      &    3.6 &   $ 5.0{\times}10^{40} $   & $4.3{\times}10^{-17}$    \\
PKS\,0232$-$04      & 1.4450   &   $=$  &  4.0     & 116    &  3670      &   97.1 &   $ 7.8{\times}10^{42} $   & $3.6{\times}10^{-16}$   \\
\hline
\multicolumn{9}{l}{$^a$Quasar systemic redshift repeated from Table \ref{table:quasars} for convenience.}\\
\multicolumn{9}{l}{$^b$Nebula detection flag with ``='' indicating a detection and ``$<$'' indicating a non-detection in which case all nebula}\\
\multicolumn{9}{l}{\  measurements represent upper limits.}\\
\multicolumn{9}{l}{$^c$FWHM of the [O\,II] surface brightness profiles, {which have systematic uncertainties of $\approx10\%$}.}\\
\multicolumn{9}{l}{$^d$Projected linear sizes of the [O\,II] nebula defined as the major axis diameter, {which have systematic uncertainties of $15\%$}.}\\
\multicolumn{9}{l}{$^e$Projected areas of the [O\,II] nebula, {which have systematic uncertainties of $\approx 20\%$}.}\\
\multicolumn{9}{l}{$^f$Total signal-to-noise ratios of the [O\,II] nebula.}\\
\multicolumn{9}{l}{$^g$Total luminosities of the [O\,II] nebula, {which have systematic uncertainties of $\approx 25\%$}.}\\
\multicolumn{9}{l}{$^h$Mean cosmological dimming corrected surface brightness measured in an annulus with inner and outer radii of $10$ and $30$}\\
\multicolumn{9}{l}{\,  { kpc, which have systematic uncertainties of $\approx 25\%$}.}\\
\multicolumn{9}{l}{$^i$ {For more details on this nebula and its environment, see \cite{Johnson:2018}.}}\\
\multicolumn{9}{l}{$^j$ {For more details on this nebula and its environment, see \cite{Zhao:2023} and \cite{Liu:2024}.}}\\
\multicolumn{9}{l}{$^k$ {For more details on these nebulae and their environments, see \cite{Chen:2023}}}\\
\multicolumn{9}{l}{$^l$ {For more details on this nebula and its environment, see \cite{Johnson:2022}.}}\\

\end{tabular}
\end{table*}

To identify large nebulae around the CUBS and MUSEQuBES quasars, we first visually inspected the quasar light and continuum-subtracted MUSE datacubes by searching for extended emission within $|\Delta v|<2000$ \kms\ of the expected wavelength of
the [O\,II] doublet. We chose this large velocity interval to be sufficient to include emission from bound gas in even massive
galaxy clusters. Furthermore, the velocity interval is sufficient to account for potential systematic uncertainty in
the quasar systemic redshifts, even for those with redshifts based on Mg\,II or broad H$\beta$. We focused the search on
the [O\,II] doublet because the MUSE spectral range covers it for all of the CUBS and MUSEQuBES quasars.
Moreover, detailed studies of individual nebulae around quasars at $z\approx 0.5-0.6$, where a more comprehensive suite of emission lines is available, suggest
 that [O\,II] dominated regions of circum-quasar nebulae are often more extended than [O\,III] dominated areas \citep[e.g.][]{Chen:2023, Liu:2024}. Our initial inspection revealed that most of the CUBS and MUSEQuBES quasars exhibit extended [O\,II] emission on $\gtrsim20-30$ kpc scales.

{To enable quantitative detection and analysis of the nebulae, we performed thresholded detection and segmentation in three dimensions
(3D; 2 spatial + 1 spectral) following the approach often used in studies of Ly$\alpha$\ nebulae at $z=2-3$ \citep[e.g.][]{Borisova:2016, Arrigoni-Battaia:2019a, Sanderson:2021}.
In particular, we formed a signal-to-noise ratio (SNR) datacube
and then smoothed it in both the spatial and spectral dimensions using a Gaussian kernel with $\sigma\approx 1.5$ pixels, though with a somewhat larger $\sigma$
in cases with elevated noise due to night sky lines. We then performed 3D source detection and segmentation on the smoothed SNR maps with a threshold of ${\rm SNR}\approx1.5$, though with a slightly higher threshold in cases where the night sky line residuals are elevated relative to the variance map.
We then connected adjacent pixels above this threshold and required a minimum of ten connected pixels and a total ${\rm SNR}\gtrsim5$ to define a detection.  To uniformly visualize and characterize these nebulae, we produced [O\,II] surface brightness maps by integrating the unsmoothed flux in each spaxel over the wavelength interval defined by the 3D segmentation. For spaxels that do not include any detected nebulae,
the background and noise properties correspond to those expected of three spectral pixels at the wavelength where the mean SNR per pixel of the nebula is highest.
The resulting [O\,II] surface brightness maps for each quasar are shown in Figure \ref{figure:SB_all}.}

{These 3D extractions have the advantage of visualizing
the full extent of detected emission while reducing noise for kinematically quiescent or undetected regions compared to surface brightness maps defined by fixed velocity intervals. However, they also exhibit complicated, spatial varying noise characteristics
\citep[see discussion in][]{Borisova:2016, Arrigoni-Battaia:2019, Mackenzie:2021}.
To quantify systematic uncertainty in the measured properties of the nebulae, we also produced surface brightness maps integrated over fixed velocity intervals chosen for each quasar field to include 95\% of the detected [O II] nebular emission in the quasar-light and continuum subtracted datacubes.
The resulting narrow-band surface brightness maps are broadly consistent with those obtained from 3D segmentation. However, the SNR in the 3D segmentation-based maps is noticeably improved compared to synthetic narrow-band images in cases where the nebular emission is spread over a wide velocity range. To ensure the robustness of the results, we performed all of
the analysis in this paper on the surface brightness maps produced by 3D segmentation and synthetic narrow bands. We confirmed that the conclusions of the paper hold with both approaches.}

{To characterize the size and flux of the nebulae around the CUBS and MUSEQuBES quasars relative to the seeing and noise level, we used the source measurement functions of Photutils \citep[][]{Bradley:2016} to measure the FWHM of the nebulae in the surface brightness maps. We integrated the 3D segmentation to estimate their total [O\,II] flux. We define extended nebulae to be detections if their FWHM exceeds the FWHM seeing measured in the datacube and if their total SNR is greater than five.
With these criteria, all but three of the quasars
have detected [O\,II] emitting, extended nebulae surrounding them. {One of these, HE\,2305$-$5315, does exhibit a small nebula offset from the quasar centroid, but its FWHM is comparable to the seeing, so we do not consider it further in this paper.}
In all 27 cases with detected, extended nebulae, the measured FWHM of the nebulae exceed the seeing FWHM by a factor of $>2.4$. This ensures that the effect of seeing on measured properties of the nebulae are minimal. For example, the contributions of the seeing to the measured FWHM of the nebulae are $<10\%$ when considered in quadrature.}

To estimate the size of the nebulae, we measured the {major axis} length-scale determined by the
maximum projected separation between pixels contained within the SNR-defined segmentation map corresponding to each nebula.
{We adopted this definition of the size of the nebulae to enable comparison with studies of $z\approx3$ Ly$\alpha$\ nebulae around quasars such as
\cite{Borisova:2016} and \cite{Mackenzie:2021} which used a similar approach.}
In two cases, PKS\,0405$-$123 \citep[][]{Johnson:2018}
and TXS\,0206$-$048 \citep[][]{Johnson:2022}, multiple large nebulae are detected in the quasar environment, some of which are well offset from
the quasars themselves. For these, we report the projected size for the largest contiguous nebula as defined in the 3D segmentation.

{Several of the nebulae are quite elongated along one axis, meaning that the major axis size does not fully encapsulate the extent of the systems.
For this reason, we also report the area in square kpc defined by the projection of the 3D segmentation of each nebula.
In cases with multiple nebulae, we report the sum of their areas.
Finally, to quantify the surface brightness of the nebulae, we computed the redshift dimming corrected surface brightness measured in an annulus
centered on the quasar with an inner radius corresponding to $10$ kpc and an outer radius corresponding to $30$ kpc, which we denote as ${\rm SB}_{10-30}(1+z)^4$.  The outer radius of $30$ kpc is chosen to be half of
the median major axis full size of the nebulae around the CUBS and MUSEQuBES quasars. The inner radius is chosen to be large enough to avoid residuals from quasar light subtraction and to exceed
the effective radii expected in stellar continuum images of late-type and early-type massive galaxies at intermediate redshift
\citep[e.g.][]{van-der-Wel:2014}.
The FWHM, major axis size, projected area, total SNR, total [O\,II] luminosity, and $\rm SB_{10-30}(1+z)^4$ measured for the nebulae are reported in Table \ref{table:nebulae}.} {To quantify the systematic uncertainty in these quantities, we remeasured them using the synthetic narrow-band images of the
[O\,II] nebulae. While the systematic differences inferred from the median ratios between the quantities measured with the 3D segmentation approach
versus the synthetic narrow-band images are minimal, there is a non-negligible scatter corresponding to 10\% for FWHM, 15\% for size, 20\% for area, and 25\% for both luminosity and ${\rm SB_{10-30}}(1+z)^4$}.

{While half of the 30 quasars in this work have HST images available in the archive, the images were acquired with long exposures to enable study of the morphologies of faint foreground galaxies and cross-correlate with CGM/IGM absorption. As a result, the bright quasars are saturated, preventing studies of stellar continuum from the host galaxies. However, 
the mean effective radius of the stellar component of massive late- and early-type galaxies at these redshifts is $R_{\rm eff}\approx9-10$ kpc \citep[e.g.][]{van-der-Wel:2014}, corresponding to a diameter of $\approx 20$ kpc. If quasar hosts represent $2\sigma$ outliers in the mass-size relation, we expect their effective radii to be $\approx15$ kpc, corresponding to a diameter of 30 kpc.}

{To gain insights into the extent of the nebulae in comparison to the expected extents of stellar components of their host galaxies, we compute the fraction of systems with nebulae larger than 100, 50, 30, and 20 kpc in the following}.
{Among the thirty quasars, seven ($23^{+9}_{-6}\%$) have detected [O\,II] nebulae with major axis sizes greater than $100$ kpc,
twenty ($67^{+7}_{-9}\%$) greater than 50 kpc, 25 ($83^{+5}_{-9}\%$) greater than 30 kpc, and 27 ($90^{+3}_{-8}\%$) greater than 20 kpc. The median size of the [O\,II] nebulae detected around the CUBS and MUSEQuBES quasars is $60$ kpc. The sizes of the [O\,II] nebulae are therefore larger than the expected
sizes of the stellar component of their host galaxies, indicating that the majority of UV luminous quasars at $z=0.4-1.4$ are surrounded by [O\,II] emitting, CGM-scale nebulae.}

Interestingly, the smaller $20-60$ kpc
nebulae are generally well centered on the quasars and relatively symmetric, but the larger ones often exhibit irregular morphology and may not be well
centered (e.g. PKS\,0405$-$123 and Q1354+048). In some cases, the morphologies of the nebulae are coincident with
nearby galaxies with redshifts similar to the quasars as previously reported for PKS\,0405$-$123 \citep{Johnson:2018}, TXS\,0206$-$048 \citep{Johnson:2022}, and HE\,0238$-$1904 \citep{Liu:2024}.
{This can be explained if the larger nebulae are often result from
ram pressure and tidal stripping experienced during galaxy interactions as previously suggested by
\cite{Stockton:1987}. The nebulae do not exhibit obvious coincidences with the radio lobe and jet orientation of the radio-loud quasars,
disfavoring jet-driven outflow origin \citep[though see][]{Fu:2009}. 
Further investigation of the coincidences between the extended nebulae and interacting galaxies in the quasar host environment will
require detailed investigation of the group environments like those in \cite{Johnson:2018, Johnson:2022, Helton:2021}, and \cite{Liu:2024}.}

\section{Discussion}
\label{section:discussion}

\begin{figure*}
\includegraphics[width=\textwidth]{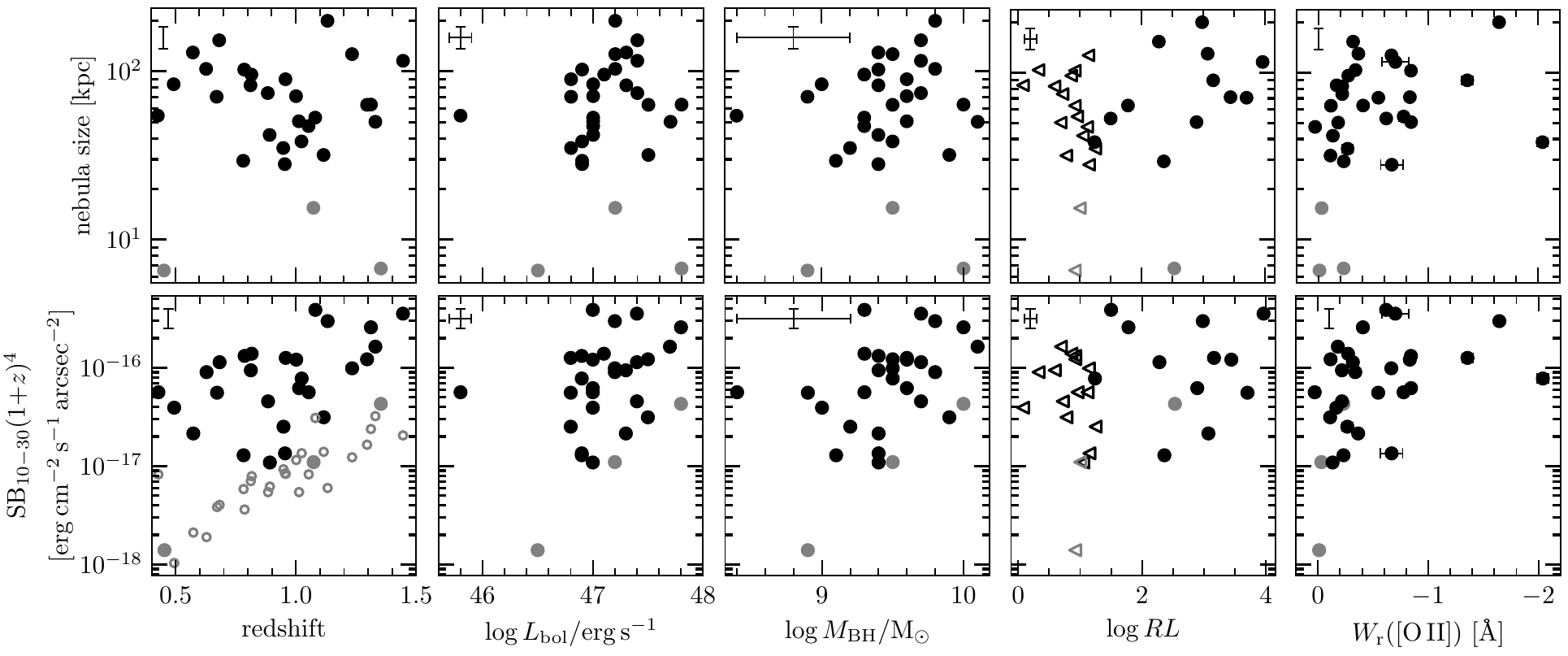}
\caption{Projected linear sizes ({\it top panels}) and {dimming corrected surface brightness measured in an annulus of $10-30$ kpc radius ({\it bottom panels}), ${\rm SB}_{10-30}(1+z)^4$},  of the [O\,II] emitting nebulae detected
around the CUBS and MUSeQuBES quasars vs quasar properties including ({\it from left to right})
redshift, bolometric luminosity, black hole mass, radio-loudness, {and rest-frame [O\,II] equivalent width}. {In all panels, quasars with detections of extended, [O\,II] emitting quasars are shown in black, while upper limits from non-detections are shown in grey.}  Quasars with non-detections in the radio data are shown
as open triangles marking the upper limit on their radio-loudness. {For reference, the $5\sigma$ surface brightness detection limit in ${\rm SB}_{10-30}(1+z)^4$ for the quasars are shown in the bottom left panel as small open symbols}.
{Systematic uncertainties are visualized with error bars in the top left of each panel. Statistical uncertainty in [O\,II] equivalent width
are shown as error bars, though most are smaller than the data points. While most of the panels exhibit significant scatter and little evidence of a correlation, the sizes and surface brightnesses
of the nebulae are tentatively correlated with rest-frame [O\,II] equivalent width with a generalized Kendall $\tau$ test returning correlation coefficients of $-0.3$ and $p$-values of $0.01$.}}
\label{figure:correlations}
\end{figure*}

Empirical characterizations of cool CGM around quasars have advanced significantly
in the last decade, thanks to a combination of large ground-based surveys like SDSS
that enable studies of Mg\,II absorption at $z\lesssim2$ \citep[e.g.][]{Bowen:2006, Farina:2014, Johnson:2015b, Chen:2023a},
dedicated absorption surveys with large telescopes at $z=2-4$ in rest UV absorption \citep[e.g.][]{Hennawi:2006, Prochaska:2013, Lau:2016},
and observations of Ly$\alpha$ emission nebulae at $z>2$ \citep[e.g.][]{Borisova:2016, Cai:2019, OSullivan:2020, Fossati:2021, Mackenzie:2021}.
The discovery of rest-frame optical emission from $>20-30$ kpc ionized nebulae around {the majority} of
UV-luminous quasars at $z=0.4-1.4$ represents a new opportunity to study the relationship between cool CGM and
the galactic systems they surround while also opening a new set of physical diagnostics based on non-resonant emission lines.

\subsection{{Possible relationships between nebulae and quasar properties}}
Previous studies of the cool, absorbing CGM around quasars at $z<2$ demonstrated that strong Mg\,II
absorbers, which typically trace cool ($T\approx 10^4$ K), high column density ($\log N({\rm H\,I})/{\rm cm^{-2}}\gtrsim 17$) enriched gas,
are common around quasar host galaxies \citep[][]{Bowen:2006, Farina:2014, Johnson:2015b}. The Mg\,II covering factors
observed around quasar hosts is significantly greater than those  found around luminous red galaxies \citep[e.g.][]{Huang:2016}
and comparable to or higher than those of massive star-forming galaxies \citep[e.g.][]{Huang:2021}.
Moreover, \cite{Johnson:2015b} identified a strong correlation between Mg\,II absorption covering fractions 
at $d<200$ kpc and quasar luminosity with luminous quasars ($\log L_{\rm bol}/{\rm erg\,s^{-1}}>45.5$) exhibiting Mg\,II covering fractions
approximately double those of lower luminosity quasars. Early studies of large-scale optical emission around quasars
based on a combination of narrow-band [O\,III] imaging and follow-up long slit spectroscopy suggested that 
emitting nebulae on scales of tens of kpc are common around radio-loud quasars and uncommon around radio-quiet ones \citep[e.g.][]{Boroson:1985, Stockton:1987}. This correlation between extended nebulosity and radio properties led to suggestions that the nebulae arise from radio-mode feedback \citep[e.g.][]{Fu:2009}.

To search for correlations between properties of the quasars and surrounding nebulae,
Figure \ref{figure:correlations} displays plots of the sizes and dimming-corrected surface brightnesses (${\rm SB_{10-30}}(1+z)^4$)
of the nebulae versus quasar redshift,
bolometric luminosity, black hole mass, radio-loudness, {and [O\,II] rest-frame equivalent width}.
The panels in Figure \ref{figure:correlations} exhibit
substantial scatter with no clear trend between quasar redshifts, bolometric luminosities,  and the sizes and areas of the [O\,II] nebulae.
{One the other hand, the [O\,II] emission levels in the quasar spectra (with negative values indicating stronger emission
relative to the continuum) appear to be weakly correlated with the sizes and surface brightnesses of the nebulae, though with substantial scatter (see the right panel of Figure \ref{figure:correlations}).}
We quantified the correlations and level of significance with the generalized Kendall $\tau$ test,
accounting for upper limits and non-detections following  \cite{Isobe:1986}. {Tests for correlations between
the properties of the nebulae and quasar redshifts, bolometric luminosities, black hole masses, and radio-loudness returned
no statistically significant results ($p$-values between 0.2 and 0.8), consistent with the lack of a visual trend seen in Figure \ref{figure:correlations}.
On the other hand, the apparent correlation between [O\,II] equivalent width and the sizes and surface brightnesses of the nebulae is marginally significant, with a correlation coefficient of $-0.3$ (indicating a positive correlation between the emission-line strengths in the quasar spectra and large-scale nebulae) and a $p$-value of 0.01 (2.5$\sigma$). To demonstrate that the extended nebulae themselves are not responsible for this correlation, we re-measured the [O\,II]
equivalent widths in quasar spectra extracted with an aperture radius of 0.6 arcsec (3-5 kpc at $z=0.4-1.4$) and found consistent results.
The tentative correlation between [O\,II] equivalent width and the nebulae suggests a general correlation
between circum-nuclear and ISM-scale gas observed in the quasar spectra and larger-scale gas supplies in the massive
halos that host UV luminous quasars (see \cite{Li:2024}). This trend is reminiscent of a previously reported potential correlation between ISM or nuclear
[O\,II] emission in luminous red galaxies and Mg\,II absorption in their CGM at projected distances of $d \lesssim 50$ kpc \citep[][]{Huang:2016}.}

The lack of correlations between the properties of the nebulae and {the bolometric luminosities} of the quasars that they surround
suggests that the presence of the emitting gas may be unrelated to {quasar outflows in many cases.}
Alternatively, the lack of correlations between the nebulae and quasar luminosity could be partly due to a lack of dynamic range in the sample.
The CUBS and MUSEQuBES surveys do not include any quasars of $\log L_{\rm bol}/{\rm erg\,s^{-1}}<45.5$.
The correlation between Mg\,II absorption at $d<200$ kpc and quasar
luminosity reported by \cite{Johnson:2015b} sets in when comparing luminous quasars of
$\log L_{\rm bol}/{\rm erg\,s^{-1}}>45.5$ with less luminous ones of $\log L_{\rm bol}/{\rm erg\,s^{-1}}<45.5$,
and this study reported no trend within the more luminous quasar population.
Furthermore, previous studies of the extended narrow-line regions around obscured AGN found a trend between
luminosity and [O\,III]-detected narrow-line region size with AGN of luminosity $\log L_{\rm bol}/{\rm erg\,s^{-1}}<45.5$ exhibiting
narrow line region sizes of $R_{\rm NLR}\lesssim 6$ kpc and more luminous obscured quasars exhibiting larger extended
narrow-line regions with $R_{\rm NLR}\approx 10-15$ kpc \citep{Sun:2017}. We note that these narrow-line region scales are 
 smaller than most of the nebulae found around the CUBS and MUSEQuBES quasars, which is likely the result of the choice of emission line ([O\,II] versus [O\,III]) or increased sensitivity of MUSE relative to more traditional long-slit spectrographs. Consequently, determining whether the correlation between cool CGM and quasar luminosity extends to ionized, line-emitting nebulae on circumgalactic scales will require samples of lower-luminosity systems observed with wide-field IFS.

The lack of correlation between the presence, size, and surface brightness of the [O\,II] emitting nebulae and
radio-loudness also stands in contrast to previous studies, which found [O\,III] emitting nebulae
around radio-loud quasars but not radio-quiet ones \citep[e.g.][]{Boroson:1985, Stockton:1987}.
While the nebulae presented in this work are observed in [O\,II] for the sake of uniformity given our lack
of [O\,III] coverage for the $z>0.85$ systems, the results are not substantially different for [O\,III] within the lower-$z$ CUBS and MUSEQuBES quasars.

Our finding of common large nebulae around radio-quiet quasars
compared to their absence in previous surveys may be the result of the sensitivity of MUSE.
The larger telescope aperture, higher throughput, and higher spectral resolution of MUSE result in 
dramatically improved surface brightness limits compared to previous long-slit spectroscopy and narrow-band imaging surveys.
Indeed, Ly$\alpha$ nebulae around $z>2$ quasars
were commonly detected around radio-loud systems but were thought to be rare around radio-quiet ones
until new, more sensitive observations with MUSE and KCWI became available. The increased sensitivity 
resulted in the discovery of ubiquitous, $100$ kpc-scale nebulae around radio-quiet quasars at high redshift,
despite non-detections with less sensitive instruments
\citep[see discussion in][]{Borisova:2016, Cantalupo:2017}.
{However, there is no correlation between surface brightness and radio loudness
within the sample of 30 quasars studied here. Reconciling this result with previous studies that found a correlation
between radio loudness and the presence of [O\,III] emitting nebulae will likely require larger samples observed with
wide-field IFS data covering both [O\,II] and [O\,III]}.

\subsection{Comparison with giant Ly$\alpha$ nebulae observed around quasars at $z=2-4$}

The discovery that half of luminous quasars at $z\approx 1$
are surrounded (in projection) by [O\,II] emitting nebulae with sizes greater than $60$ kpc indicates
that dense, cool CGM is common in quasar-host halos at this epoch. However, this represents a significantly
lower incidence and smaller size-scale than findings of ubiquitous, $\gtrsim100$ kpc H\,I Ly$\alpha$ emitting nebulae around
quasars at $z>2$
\citep[][]{Borisova:2016, Cai:2019, OSullivan:2020, Fossati:2021, Mackenzie:2021}. However, physical interpretation of this
apparent difference requires accounting for the expected [O\,II] to H\,I Ly$\alpha$ line ratio and the difference
in surface brightness dimming between $z\approx 1$ and $z>2$.

\begin{figure}
\includegraphics[width=\columnwidth]{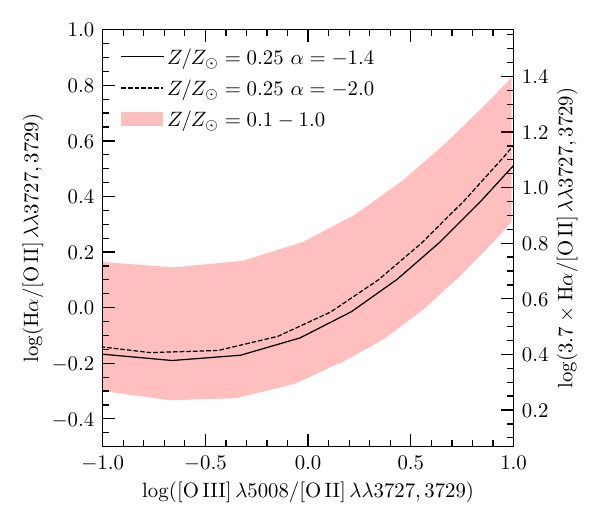}
\caption{Expected line ratios of H$\alpha$/[O\,II] as a function of [O\,III]/[O\,II] ratio for dust-free AGN photoionized gas  of {$Z=0.25Z_\odot$ (black lines) and with metallicities of $Z=0.1-1.0 Z_\odot$ (faded red band)}.
The gas metallicity is chosen to be consistent with typical metallicities of absorbing CGM around quasars at $z=2-4$ \citep{Lau:2016}
and luminous red galaxies at $z=0.4$ \citep[][]{Zahedy:2019}. The two lines show the predicted line ratio for ionizing spectrum power-law slopes of {$\alpha=-1.4$ and $-2.0$ in solid and dashed black line for reference}.
The right axis is scaled to $3.7\times$H$\alpha$/[O\,II] to approximate the expected $\rm Ly\alpha$/[O\,II] ratio under the assumption that
the Ly$\alpha$ emission is {primarily from recombination with modest scattering} \citep[see discussion in ][]{Langen:2023}. Observations of ionized nebulae around quasars at $z=0.4-0.7$ where the [O\,III] to [O\,II] ratio is constrained with optical data indicate that the line ratios range from $\log \rm [O\,III]/[O\,II]=-1$ to $1$.}
\label{figure:ratios}
\end{figure}

The radiative transfer involved in modeling Ly$\alpha$ is complex, so we opt for an empirical approach and adopt a measured
$\rm Ly\alpha/H\alpha$ ratio observed around quasars and then combine this with model expectations for $\rm H\alpha/[O\,II]$ 
to determine whether we expect the types of nebulae observed in Ly$\alpha$ at $z=2-4$ around quasars to be observable in
[O\,II] at $z\approx 1$. 
{To set expectations for $\rm H\alpha/[O\,II]$, we considered emission models of moderately enriched, dust-free gas photoionized by an AGN
{using Cloudy v17.03 \citep[][]{Ferland:2017} with model parameters motivated by \cite{Groves:2004}}. 
{In particular, we chose a power-law ionizing spectrum with spectral slope ranging from $\alpha=-2.0$ to $-1.4$.}
{We note that the range of predicted H$\alpha$/[O\,II] ratios for AGN photoionized gas are similar to predictions
for fast shocks \citep[e.g.][]{Allen:2008} and star-forming galaxies \citep[e.g.][]{Kewley:2004}, indicating that our conclusions
are not sensitive to the ionization source.}

{One of the dominant drivers of variation in expected H$\alpha$/[O\,II] ratios is gas metallicity.}
While the metallicity of the CGM around quasars at $z\approx 1$ is not well studied due to observational limitations,
 \cite{Liu:2014} inferred approximately solar metallicity in the nebula around HE\,0238$-$1904 for regions detectable in weak-line diagnostics.
 Furthermore, the cool absorbing CGM around luminous quasars at $z=2-4$ from \cite{Lau:2016} and luminous red galaxies (LRGs) at $z=0.4$
from \cite{Zahedy:2019} exhibit typical metallicities of $Z\approx0.25\,Z_\odot$, though with substantial scatter in the case of LRGs.
{We therefore adopt a fiducial metallicity of $Z=0.25\, Z_\odot$ for the AGN photoionized gas models but also explore a range of $Z=0.1-1.0Z_\odot$.} The presence of substantial dust
in large nebulae around quasars is disfavored in cases where Balmer line ratios are measurable \citep[e.g.,][]{Helton:2021, Liu:2024}.}
The left axis of Figure \ref{figure:ratios} shows the $\rm H\alpha/[O\,II]$  ratio versus  $\rm [O\,III]/[O\,II]$
over the range of  $\rm [O\,III]/[O\,II]$ observed in extended nebulae around $z\approx 0.6$ quasars \citep[][]{Johnson:2018, Helton:2021, Liu:2024}.

 To translate this expected  $\rm H\alpha/[O\,II]$ into expected $\rm Ly\alpha/[O\,II]$, we multiply by a factor of 3.7
corresponding to the observed $\rm Ly\alpha/H\alpha$ ratio of {$3.7\pm 0.3$} observed in nebulae around {three} $z\approx3$ quasars from \cite{Langen:2023}.
{We note that ratio is a factor of $\approx2-3$ below that expected from pure recombination radiation. \cite{Langen:2023}
and previous works \citep[e.g.][]{Cantalupo:2019} attribute this decreased $\rm Ly\alpha/H\alpha$ to radiative transfer effects combined with aperture losses.} Combining this empirical $\rm Ly\alpha/[O\,II]$ ratio with the modeled $\rm H\alpha/[O\,II]$ we expect $\log ({\rm Ly\alpha/[O\,II]})\approx 0.25-1.4$ assuming
AGN photoionization, $\rm Ly\alpha$ emission {arising primarily from recombination with moderate scattering}, and modest metal enrichment levels. The expected  $\rm Ly\alpha/[O\,II]$ ratios are shown as a function of $\rm [O\,III]/[O\,II]$ in the right axis of Figure \ref{figure:ratios}. {The presence of dust would only increase the strength of [O\,II] relative to Ly$\alpha$.}

\begin{figure*}
\includegraphics[width=\textwidth]{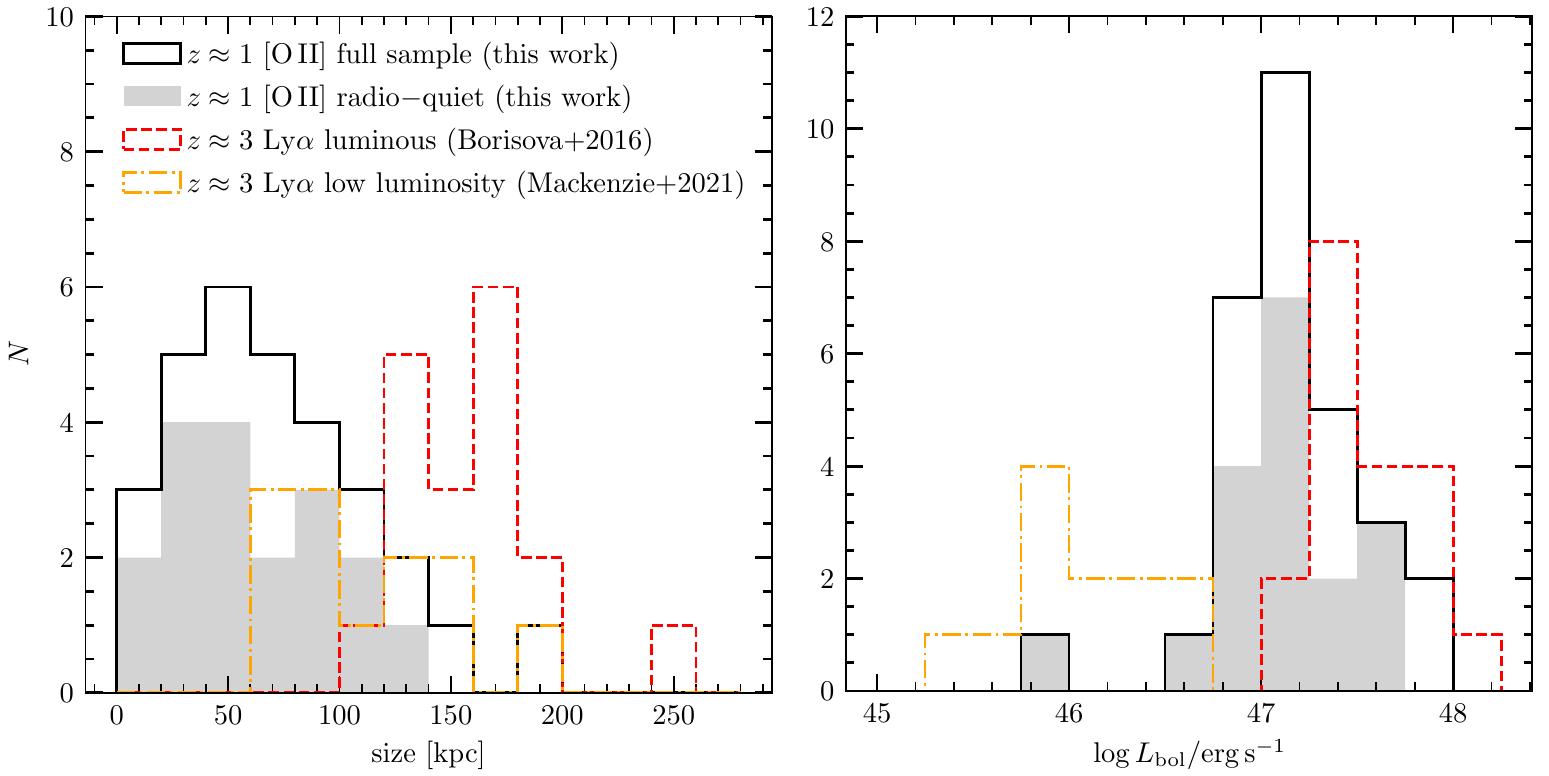}
\caption{{{\it left panel}: Comparison of the sizes of [O\,II] emitting nebulae discovered around CUBS and MUSEQuBES quasars at $z=0.4-1.4$ from this work with Ly$\alpha$ nebulae around $z\approx 3$ quasars \cite[][]{Borisova:2016, Mackenzie:2021}. The full CUBS and MUSEQuBES [O\,II] sample is shown in black, while the sub-sample of radio-quiet systems is shown in solid grey histogram.
For the $z\approx3$ Ly$\alpha$ nebulae, the size distribution of luminous quasars from \protect \cite{Borisova:2016} is shown in red dashed line while that of less luminous quasars from \protect \cite{Mackenzie:2021} is shown in orange dash-dotted line. The [O\,II] nebulae around $z=0.4-1.4$ quasars are substantially smaller than Ly$\alpha$ nebulae at $z\approx 3$ regardless of radio properties. {\it right panel}: Comparison of the luminosities of the CUBS and MUSEQuBES quasars with the $z=3$ samples from \cite{Borisova:2016} and \cite{Mackenzie:2021}. The [O\,II] nebulae discovered around the CUBS and MUSEQuBES quasars are smaller and more rare than the Ly$\alpha$ nebulae discovered around $z\approx3$ quasars \cite{Borisova:2016} and low-luminosity quasars  \cite{Mackenzie:2021} even though the luminosities of the CUBS and MUSEQuBES quasars fall within the range spanned by the two $z=3$ samples.}}
\label{figure:sizes}
\end{figure*}

Surveys of $\rm Ly\alpha$ emission around quasars at {$\langle z\rangle \approx3.3$} such as \cite{Borisova:2016} and \cite{Mackenzie:2021} achieve observed-frame surface brightness limits
similar to the ones enabled by the CUBS and MUSEQuBES observations. However, the difference in surface brightness dimming at the {mean redshift of the 
\cite{Borisova:2016} and \cite{Mackenzie:2021} studies} versus
{$\langle z \rangle =0.95$ in this work introduces a factor of $(1+3.3)^4/(1 + 0.95)^4\approx24$ difference in rest-frame surface brightness sensitivity. This factor of 24 difference in surface brightness dimming is equivalent to $1.4$ dex, compensating 
for the expectation that [O\,II] will be a weaker line
than Ly$\alpha$.} 
Therefore, if the cool, $\rm Ly\alpha$ emitting CGM observed  around quasars primarily arise from recombination radiation with only modest scattering, if the quantities of cool, emitting CGM around quasars do not evolve significantly between $z\approx 3$ and $z\approx 1$, if CGM metallicities around $z\approx1$ quasars are not substantially lower than expected from absorption surveys, and if quasar episodic lifetimes are not smaller at intermediate redshift compared to $z\approx3$, we expect that the [O\,II]
nebulae observed in this work would be similar in size and incidence rate to  $\rm Ly\alpha$  at higher redshift.

In contrast to this expectation, the [O\,II] emitting nebulae found around CUBS and MUSEQuBES quasars are smaller and less common than
their Ly$\alpha$ counterparts at $z\approx 3$ as shown in the left panel Figure \ref{figure:sizes}. In particular, while $100$ kpc-scale Ly$\alpha$ nebulae
are fairly ubiquitous at $z=3$, only seven of the thirty ($23^{+9}_{-6}\%$) of the CUBS and MUSEQuBES quasars exhibit such large [O\,II] nebulae.
{The smaller sizes of the [O\,II] nebulae at $z\approx1$ are even starker when considered relative to the estimated virial radii
of the quasar host halos, which are $\approx 120$ kpc at $z\approx3$ \citep[][]{Shen:2007} compared to $\approx500$ kpc for UV luminous quasars at $z\approx1$ \citep[][]{Li:2024}.}
The fact that [O\,II]
emitting nebulae are smaller and less common around $z\approx 1$ quasars than Ly$\alpha$ nebulae around $z=3$ ones can be explained if a substantial
fraction Ly$\alpha$ in the outer regions of the $z=3$ nebulae arise from scattered Ly$\alpha$ photons {in a more neutral phase of the CGM that does not emit in [O\,II]}. The contrast could also be explained if there is substantial evolution in the {quantities or density distribution} of the cool, dense phase of halo gas around quasars between $z\approx3$ and $z\approx 1$
or if the metallicities of the CGM of $z\approx1$ quasar hosts on $\approx 50$ kpc scales are substantially lower than $0.25 Z_\odot$.
We note, however, that the $Z\approx 0.25 Z_\odot$ metallicity used to estimate the [O\,II] to Ly$\alpha$\ ratio in Figure \ref{figure:ratios} is based on absorption measurements in massive halos at distances of $d\approx50-100$ kpc \citep[e.g.,][]{Lau:2016, Zahedy:2019}, similar to the size-scale of the Ly$\alpha$ nebulae.
Further, \cite{Johnson:2015b} found little evidence
for evolution in Mg\,II covering factor observed in the CGM of quasars over this same cosmic period. 
{Alternatively, the difference in size and incidence of Ly$\alpha$ nebulae at $z\approx 3$ and
[O\,II] emitting nebulae at $z\approx1$ could also be explained if luminous quasar episodic lifetimes at $z\approx1$ are shorter than at $z\approx 2-3$.}
Further investigation of the differences in optically-emitting [O\,II] nebulae vs UV emitting Ly$\alpha$ nebulae around quasars requires multi-wavelength observations at both redshifts.

\section{Summary and conclusions}
\label{section:conclusions}

In this paper, we presented the first comprehensive search for non-resonant, optically emitting [O\,II] nebulae around $30$ UV luminous quasars
at $z=0.4-1.4$ using deep integral field spectroscopy acquired as part of the Cosmic Ultraviolet Baryons Survey and MUSE Quasar Blind Emitters
Survey. Our findings indicate that $>60$ kpc-scale circumgalactic nebulae are common around UV luminous quasars, with approximately half of
the sample exhibiting such large-scale emission while {$\approx 80-90\%$ exhibit emission on $>20-30$ kpc scales}. Within the CUBS and MUSEQuBES quasar samples,
the presence and size of the [O\,II] nebulae are not correlated with quasar redshift, luminosity, supermassive black hole mass, or radio-loudness.
{On the other hand, the sizes and surface brightnesses of the [O\,II] nebulae are tentatively correlated ($2.5\sigma$ significance) with the rest-frame [O\,II] equivalent width detected in the quasar spectra due to circumnuclear and ISM-scale ionized gas.}
Further investigation of the relationship between quasars and their surrounding cool gas supplies will require larger samples extending over a broader range of quasar properties, particularly to lower luminosities and black hole masses.

The [O\,II] emitting nebulae around quasars at $z\approx 1$ found in this work
are less common and smaller on average than Ly$\alpha$ emitting nebulae found around quasars at $z>2$. This difference can be explained, for example,
if the dense, cool phase of the CGM around quasars evolves significantly with redshift, if the metallicities in quasar-host CGM on $\approx50$ kpc scales are lower than expected from absorption surveys, if quasar episodic lifetimes are shorter at $z\approx 1$ compared to $z\approx 3$, or if the scattering of Ly$\alpha$ photons off of more neutral gas contributes
non-negligibly to the outer regions of Ly$\alpha$ halos around quasars at $z=2-3$.
These scenarios can be tested by observing the nebulae with broader spectral coverage
using upcoming near-IR IFS such as MIRMOS \citep[][]{Konidaris:2020}  and HARMONI \citep[][]{Thatte:2021}, or with new UV space-based instrumentation targeting low-$z$ systems in H\,I Ly$\alpha$.

\section*{Acknowledgements}
We thank the anonymous referee for constructive comments and suggestions that significantly strengthened this paper.
We are grateful to Eric Bell and Kayhan G\"ultekin for insightful discussions which helped shape interpretation of the results.
SDJ acknowledges partial support from HST-GO-15280.009-A, HST-GO-15298.007-A, HST-GO-15655.018-A, and HST-GO-15935.021-A. JIL is supported by the Eric and Wendy Schmidt AI in Science Postdoctoral Fellowship, a Schmidt Futures program.
Based on observations from the European Organization for Astronomical Research in the Southern Hemisphere under ESO.
The paper made use of the NASA/IPAC Extragalactic Database, the NASA Astrophysics Data System, Astropy \citep[][]{Astropy-Collaboration:2018}, and Aplpy \citep[][]{Robitaille:2012}. 

\section*{Data Availability}
The raw MUSE data, associated calibrations, and ESO reduced datacubes are publicly available
in the ESO archive.

\bibliographystyle{aasjournal}
\bibliography{astro.bib} 

\end{document}